\documentclass[conference]{IEEEtran}
\usepackage{amssymb,rotating,graphicx,cite,subeqn,calc,color,multirow}
\usepackage{graphicx,cite,calc,dsfont}
\usepackage[defs]{ams}
\usepackage{eqnarray,subeqn,xcolor}

\newtheorem{example}{\bf Example}

\newcommand\TCS{{\em IEEE Transactions on Communications,} vol. }
\newcommand\JSAC{{\em IEEE Journal on Selected Areas in Communications,} vol. }

\begin{document}

\title{Interference Cancelation in Coherent CDMA Systems Using Parallel
Iterative Algorithms}

\author{\IEEEauthorblockN{Kamal Shahtalebi}
\IEEEauthorblockA{Department of Information Technology\\
The University of Isfahan, Isfahan\\ Iran, Postal Code:
81746-73441 \\
Email: shahtalebi@eng.ui.ac.ir} \and \IEEEauthorblockN{Gholam Reza
Bakhshi} \IEEEauthorblockA{Department of Electrical and Computer\\
Engineering, Yazd University, Yazd, Iran\\
Email: farbakhshi@yahoo.com} \and \IEEEauthorblockN{Hamidreza
Saligheh Rad}
\IEEEauthorblockA{School of Engineering and Applied\\
Sciences, Harvard University\\ Cambridge MA, 02138, USA\\
Email: hamid@seas.harvard.edu}}

\maketitle

\begin{abstract}
Least mean square-partial parallel interference cancelation
(LMS-PPIC) is a partial interference cancelation using adaptive
multistage structure in which the normalized least mean square
(NLMS) adaptive algorithm is engaged to obtain the cancelation
weights. The performance of the NLMS algorithm is mostly dependent
to its step-size. A fixed and non-optimized step-size causes the
propagation of error from one stage to the next one. When all user
channels are balanced, the unit magnitude is the principal property
of the cancelation weight elements. Based on this fact and using a
set of NLMS algorithms with different step-sizes, the parallel
LMS-PPIC (PLMS-PPIC) method is proposed. In each iteration of the
algorithm, the parameter estimate of the NLMS algorithm is chosen to
match the elements' magnitudes of the cancelation weight estimate
with unity. Simulation results are given to compare the performance
of our method with the LMS-PPIC algorithm in three cases: balanced
channel, unbalanced channel and time varying channel.
\end{abstract}

\section{Introduction}\label{S1}%
Multiuser detectors for code division multiple access (CDMA)
receivers are effective techniques to eliminate the multiple access
interference (MAI). In CDMA systems, all users receive the whole
transmitted signals concurrently that are recognized by their
specific pseudo noise (PN) sequences. In such a system, there exists
a limit for the number of users that are able to simultaneously
communicate. This limitation is because of the MAI generated by
other users (see e.g. \cite{R1,R4}). High quality detectors improve
the capacity of these systems \cite{R1,Verdu862}. However their
computational complexities grow exponentially with increasing the
number of users and the length of the transmitted sequence
\cite{R6}.

Multiple stage subtractive interference cancelation is a suboptimal
solution with reduced computational complexity. In this method and
before making data decisions, the estimated interference from other
users are removed from the specific user's received signal. The
cancelation can be carried out either in a serial way (successively)
(see e.g. \cite{Viterbi90,Patel94}) or in a parallel manner (see
e.g. \cite{R2,R4,R5}). The parallel interference cancelation (PIC)
is a low computational complex method that causes less decision
delay compared to the successive detection and is much simpler in
implementation.

Usually at the first stage of interference cancelation in a multiple
stage system, the interfering data for each user which is made by
other users is unknown. PIC is implemented to estimate this data
stage by stage. In fact when MAI is estimated for each user, the bit
decision at the $(s-1)^{\rm th}$ stage of cancelation are used for
bit detection at the $s^{\rm th}$ stage. Apparently, the more
accurate the estimates are, the better performance of the detector
is. However, in the conventional multistage PIC \cite{R2}, a wrong
decision in one stage can increase the interference. Based on
minimizing the mean square error between the received signal and its
estimate from the previous stage, G. Xue \emph{et al.} proposed the
least mean square-partial parallel interference cancelation
(LMS-PPIC) method \cite{R5,RR5}. In LMS-PPIC, a weighted value of
MAI of other users is subtracted before making the decision of a
specific user. The least mean square (LMS) optimization and the
normalized least mean square (NLMS) algorithm \cite{Haykin96} shape
the structure of the LMS-PPIC method of the weight estimation of
each cancelation stage. However, the performance of the NLMS
algorithm is mostly dependent on its step-size. Although a large
step-size results in a faster convergence rate, but it causes a
large maladjustment. On the other hand, with a very small step-size,
the algorithm almost keeps its initial values and can not estimate
the true cancelation weights. In the LMS-PPIC method, both of these
cases cause propagation of error from one stage to another. In
LMS-PPIC, the $m^{\rm th}$ element of the weight vector in each
stage is the true transmitted binary value of the $m^{\rm th}$ user
divided by its hard estimate value from the previous stage. Hence
the magnitude of all weight elements in all stages are equal to
unity. This is a valuable information that can be used to improve
the performance of the LMS-PPIC method. In this paper, we propose
parallel LMS-PPIC (PLMS-PPIC) method by using a set of NLMS
algorithms with different step-sizes. The step-size of each
algorithm is chosen from a sharp range \cite{sg2005} that guarantees
stable operation. While in this paper we assume coherent
transmission, the non-coherent scenario is investigated in
\cite{noncohpaper}.

The rest of this paper is organized as follows: In section \ref{S2},
the LMS-PPIC \cite{R5} is reviewed. The LMS-PPIC algorithm is an
important example of multistage parallel interference cancelation
methods. In section \ref{S3}, the PLMS-PPIC method is explained. In
section \ref{S5} some simulation examples are given to compare the
results of PLMS-PPIC with those of LMS-PPIC. Finally, the paper is
concluded in section \ref{S6}.

\section{Multistage Parallel Interference Cancelation: LMS-PPIC Method}\label{S2}%

We assume $M$ users synchronously send their symbols
$\alpha_1,\alpha_2,\cdots,\alpha_M$ via a base-band CDMA
transmission system where $\alpha_m\in\{-1,1\}$. The $m^{th}$ user
has its own code $p_m(.)$ of length $N$, where $p_m(n)\in \{-1,1\}$,
for all $n$. It means that for each symbol $N$ bits are transmitted
by each user and the processing gain is equal to $N$. At the
receiver we assume that perfect power control scheme is applied.
Without loss of generality, we also assume that the power gains of
all channels are equal to unity and users' channels do not change
during each symbol transmission (it can change from one symbol
transmission to the next one) and the channel phase $\phi_m$ of
$m^{th}$ user is known for all $m=1,2,\cdots,M$ (see
\cite{noncohpaper} for non-coherent transmission). We define
\begin{equation}
\label{tte1} c_m(n)=e^{j\phi_m}p_m(n).
\end{equation}
According to the above assumptions, the received signal is
\begin{equation}
\label{e1} r(n)=\sum\limits_{m=1}^{M}\alpha_m
c_m(n)+v(n),~~~~n=1,2,\cdots,N,
\end{equation}
where $v(n)$ is additive white Gaussian noise with zero mean and
variance $\sigma^2$. In order to make a new variable set
$\alpha^{s}_1,\alpha^{s}_2,\cdots,\alpha^{s}_M$ for the current
stage $s$, multistage parallel interference cancelation method uses
$\alpha^{s-1}_1,\alpha^{s-1}_2,\cdots,\alpha^{s-1}_M$ (the bit
estimate outputs of the previous stage $s-1$) to estimate the
related MAI of each user, to subtract it from the received signal
$r(n)$ and to make a new decision on each user variable
individually. The output of the last stage is considered as the
final estimate of the transmitted bits. In the following we explain
the structure of the LMS-PIC method. Assume
$\alpha_m^{(s-1)}\in\{-1,1\}$ is a given estimate of $\alpha_m$ from
stage $s-1$. Let us define
\begin{equation}
\label{e6} w^s_{m}=\frac{\alpha_m}{\alpha_m^{(s-1)}}.
\end{equation}
From (\ref{e1}) and (\ref{e6}) we have
\begin{equation}
\label{e7} r(n)=\sum\limits_{m=1}^{M}w^s_m\alpha^{(s-1)}_m
c_m(n)+v(n).
\end{equation}
Define
\begin{subequations}
\begin{eqnarray}
\label{e8} W^s&=&[w^s_{1},w^s_{2},\cdots,w^s_{M}]^T,\\
\label{e9}
\!\!\!\!\!\!\!\!\!\!\!\!\!\!\!\!\!\!\!\!X^{s}(n)\!\!\!&=&\!\!\![\alpha^{(s-1)}_1c_1(n),\alpha^{(s-1)}_2c_2(n),\cdots,\alpha^{(s-1)}_Mc_M(n)]^T.
\end{eqnarray}
\end{subequations}
where $T$ stands for transposition. From equations (\ref{e7}),
(\ref{e8}) and (\ref{e9}), we have
\begin{equation}
\label{e10} r(n)=W^{s^T}X^{s}(n)+v(n).
\end{equation}
Given the observations $\{r(n),X^{s}(n)\}^{N}_{n=1}$, an adaptive
algorithm can be used to compute
\begin{equation}
\label{te1} W^{s}(N)=[w^{s}_1(N),w^{s}_2(N),\cdots,w^{s}_M(N)]^T,
\end{equation}
which is an estimate of $W^s$ after $N$ iterations. Then
$\alpha^s_m$, the estimate of $\alpha_m$ at stage $s$, is given by
\begin{equation}
\label{e11}
\alpha^{s}_m=\mbox{sign}\left(\mbox{Re}\left\{\sum\limits_{n=1}^{N}q^s_m(n)c^*_m(n)\right\}\right),
\end{equation}
where $(.)^*$ stands for complex conjugation and
\begin{equation}
\label{e12} q^{s}_{m}(n)=r(n)-\sum\limits_{m^{'}=1,m^{'}\ne
m}^{M}w^{s}_{m^{'}}(N)\alpha^{(s-1)}_{m^{'}} c_{m^{'}}(n).
\end{equation}
The inputs of the first stage $\{\alpha^{0}_m\}_{m=1}^M$ (needed for
computing $X^1(n)$) is given by the conventional bit detection
\begin{equation}
\label{e5}
\alpha^{0}_m=\mbox{sign}\left(\mbox{Re}\left\{\sum\limits_{n=1}^{N}
r(n)c^*_m(n)\right\}\right),~~~m=1,2,\cdots,M.
\end{equation}
Given the available information $\{r(n),X^{s}(n)\}^{N}_{n=1}$ and
using equation (\ref{e10}), there are a variety of choices for
parameter estimation. In LMS-PPIC, the NLMS algorithm is used to
compute $W^{s}(N)$. Table \ref{tab1} shows the full structure of the
LMS-PPIC method.

To improve the performance of the LMS-PPIC method, in the next
section we propose a modified version of it. In our method a set of
individual NLMS algorithms with different step-sizes are used.

\section{Multistage Parallel Interference Cancelation: PLMS-PPIC Method}\label{S3}%

The NLMS (with fixed step-size) converges only in the mean sense. In
the literature, $\mu\in (0,2)$ guarantees the mean convergence of
the NLMS algorithm \cite{Haykin96,R7}. Based on Cram\'er-Rao bound,
a sharper range was given in \cite{sg2005} as follows
\begin{equation}
\label{tte2} \mu\in \Psi=\left(0,1-\sqrt{\frac{M-1}{M}}\right],
\end{equation}
where $M$ is the length of the parameter under estimate. Here $M$ is
the number of users or equivalently the system load. As equation
(\ref{tte2}) shows, the range of the step-size is a decreasing
function of the system load. It means that as the number of users
increases, the step-size must be decreased and vice versa. In the
proposed PLMS-PPIC method, $\Psi$ has a critical role.

From (\ref{e6}), we have
\begin{equation}
\label{e13} |w^s_{m}|=1 ~~~m=1,2,\cdots,M,
\end{equation}
which is equivalent to
\begin{equation}
\label{e14} \sum\limits_{m=1}^{M}||w^s_{m}|-1|=0.
\end{equation}
To improve the performance of the NLMS algorithm, at time
iteration $n$, we can determine the step size $\mu(n)$ from
$\Psi$, in such a way that $\sum\limits_{m=1}^{M}||w^s_{m}(n)|-1|$
is minimized, i.e.
\begin{equation}
\label{e15} \mu(n)=\arg\min\limits_{\mu\in
\Psi}\left\{\sum\limits_{m=1}^{M}||w^s_{m}(n)|-1|\right\},
\end{equation}
where $w^{s}_m(n)$, the $m^{\rm th}$ element of $W^s(n)$, is given
by (see Table~\ref{tab1})
\begin{equation}
\label{e16} w^{s}_m(n)=w^{s}_m(n-1)+\mu
\frac{\alpha^{(s-1)}c^*_m(n)}{\|X^s(n)\|^2}e(n).
\end{equation}
The complexity to determine $\mu(n)$ from (\ref{e15}) is high,
especially for large values of $M$. Instead we propose the following
method.

We divide $\Psi$ into $L$ subintervals and consider $L$ individual
step-sizes $\Theta=\{\mu_1,\mu_2,\cdots,\mu_L\}$, where
$\mu_1=\frac{1-\sqrt{\frac{M-1}{M}}}{L}, \mu_2=2\mu_1,\cdots$, and
$\mu_L=L\mu_1$. In each stage, $L$ individual NLMS algorithms are
executed ($\mu_l$ is the step-size of the $l^{th}$ algorithm). In
stage $s$ and at iteration $n$, if
$W^{s}_k(n)=[w^s_{1,k},\cdots,w^s_{M,k}]^T$, the parameter estimate
of the $k^{\rm th}$ algorithm minimized our criteria, i.e.
\begin{equation}
\label{e17} W^s_k(n)=\arg\min\limits_{W^s_l(n)\in I_{W^s}
}\left\{\sum\limits_{m=1}^{M}||w^s_{m,l}(n)|-1|\right\},
\end{equation}
where $W^{s}_l(n)=W^{s}(n-1)+\mu_l \frac{X^s(n)}{\|X^s(n)\|^2}e(n),
~~~ l=1,2,\cdots,k,\cdots,L-1,L$ and
$I_{W^s}=\{W^s_1(n),\cdots,W^s_L(n)\}$, then it is considered as the
parameter estimate at time iteration $n$, i.e. $W^s(n)=W^s_k(n)$ and
all other algorithms replace their weight estimates by $W^{s}_k(n)$.
Table~\ref{tab2} shows the details of the PLMS-PPIC method. As
Table~\ref{tab2} shows, in stage $s$ and at time iteration $N$ where
$W^s(N)$ is computed, the PLMS-PPIC method computes $\alpha^s_m$
from equation (\ref{e11}). This is similar to the LMS-PPIC method.
Here the PLMS-PPIC and the LMS-PPIC methods are compared with each
other.
\begin{itemize}
\item Computing $\mu_{l}Z(n)=\mu_l \frac{X^s(n)}{\|X^s(n)\|^2}$,
$L$ times more than LMS-PPIC, and computing
$\sum\limits_{m=1}^{M}||w^s_{m,l}(n)|-1|$ in each iteration of each
stage of PLMS-PPIC, is the difference between it and the LMS-PPIC
method. \item Because the step-sizes of all individual NLMS
algorithms of the proposed method are given from a stable operation
range, all of them converge fast or slowly. Hence the PLMS-PPIC is a
stable method.
\item As we expected and our simulations show, choosing the
step-size as a decreasing function of system loads (based on
relation (\ref{tte2})) improves the performance of both NLMS
algorithm in LMS-PPIC and parallel NLMS algorithms in PLMS-PPIC
methods in such a way that there is no need for the third stage,
i.e. both the LMS-PPIC and PLMS-PPIC methods get the optimum weights
in the second stage. However only when the channel is time varying,
the third stage is needed, e.g. \ref{ex4}. \item Increasing the
number of parallel NLMS algorithms $L$ in PLMS-PPIC method increases
the complexity, while it improves the performance as well.
\item As our simulations show, the LMS-PPIC method is more sensitive
to the Channel loss, near-far problem or unbalanced channel gain
compared to the PLMS-PPIC.
\end{itemize}
In the following section, some examples are given to illustrate the
effectiveness of our proposed methods.

\section{Simulations}\label{S5}%

In this section we have considered some simulation examples.
Examples \ref{ex2}-\ref{ex4} compare the conventional, the LMS-PPIC
and the PLMS-PPIC methods in three cases: balanced channels,
unbalanced channels and time varying channels. Example \ref{ex2} is
given to compare LMS-PPIC and PLMS-PPIC in the case of balanced
channels.

\begin{figure} \centering
\includegraphics[width=.4\textwidth]{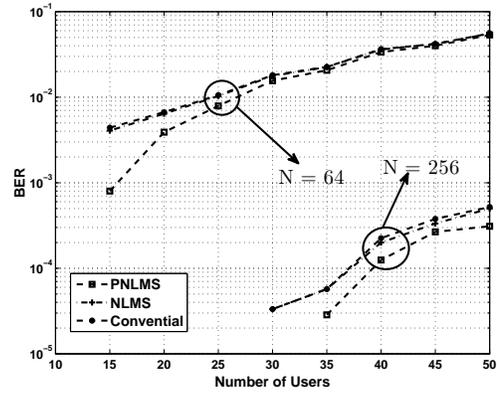}
\caption{The BER of the conventional, the LMS-PPIC, and the
PLMS-PPIC methods versus the system load in balanced channel, using
two stages for $N=64$ and $N=256$.}\label{Figex2}
\end{figure}

\begin{example}{\it Balanced channels}:
\label{ex2} Consider the system model (\ref{e7}) in which $M$ users,
each having their own codes of length $N$, send their own bits
synchronously to the receiver and through their channels. The signal
to noise ratio (SNR) is $0$dB. In this example we assume that there
is no power-unbalanced or channel loss. The step-size of the NLMS
algorithm in LMS-PPIC method is $\mu=0.1(1-\sqrt{\frac{M-1}{M}})$
and the set of step-sizes of the parallel NLMS algorithms in
PLMS-PPIC method is
$\Theta=\{0.01,0.05,0.1,0.2,\cdots,1\}(1-\sqrt{\frac{M-1}{M}})$,
i.e. $\mu_1=0.01(1-\sqrt{\frac{M-1}{M}}),\cdots,
\mu_4=0.2(1-\sqrt{\frac{M-1}{M}}),\cdots,
\mu_{12}=(1-\sqrt{\frac{M-1}{M}})$. Figure~\ref{Figex2} shows the
average bit error rate (BER) over all users versus $M$, using two
stages when $N=64$ and $N=256$. As it is shown, while there is no
remarkable performance difference between all three methods for
$N=64$, the PLMS-PPIC outperforms the conventional and the LMS-PPIC
methods for $N=256$. Simulations also show that there is no
remarkable difference between results in two stage and three stage
scenarios.
\end{example}

Although LMS-PPIC and PLMS-PPIC are structured based on the
assumption of no near-far problem, these methods (especially the
second one) have remarkable performance in the cases of unbalanced
and/or time varying channels. These facts are shown in the two
upcoming examples.

\begin{figure} \centering
\includegraphics[width=.4\textwidth]{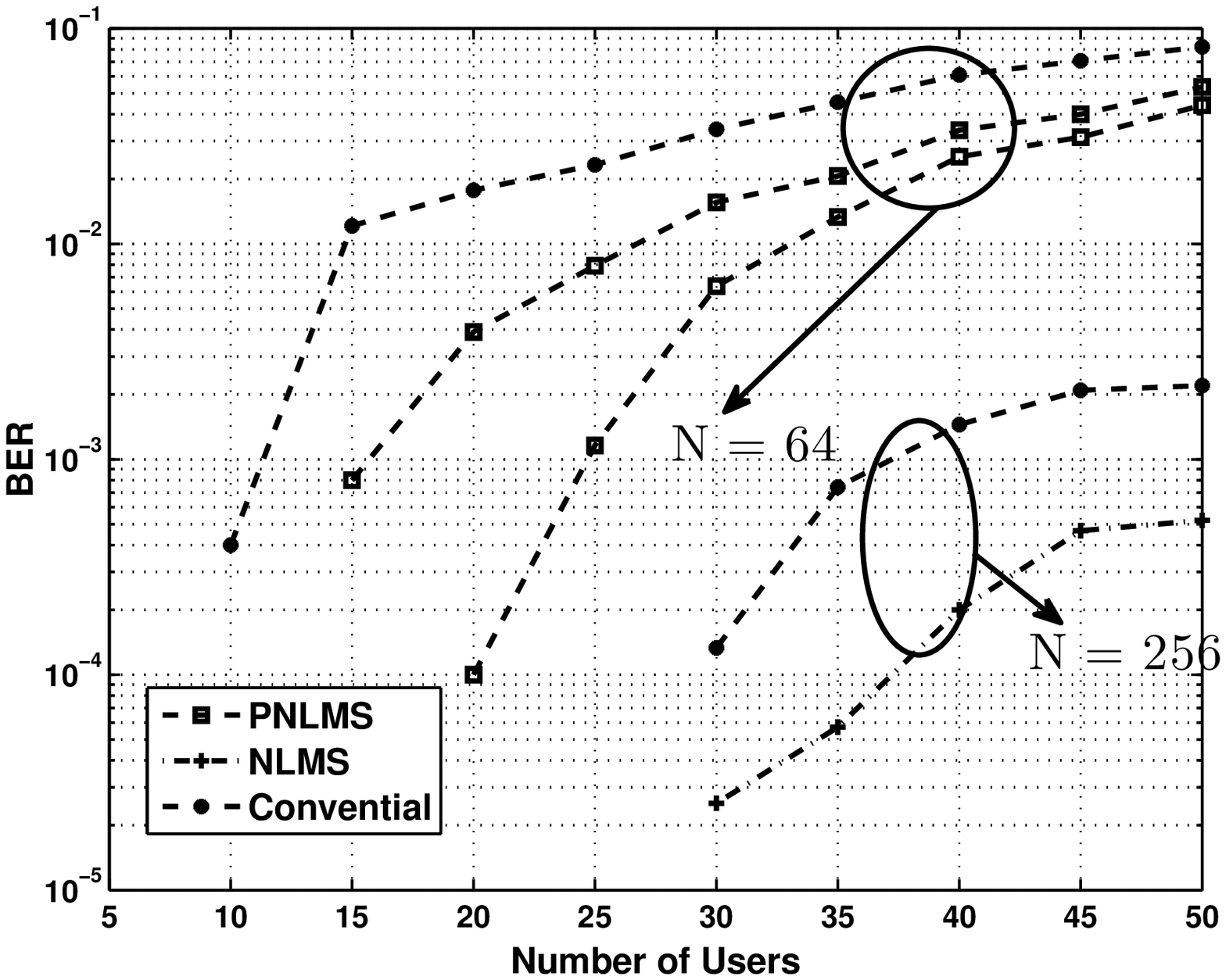}
\caption{The BER of the conventional, the LMS-PPIC, and the
PLMS-PPIC methods versus the system load in unbalanced channel,
using two stages for $N=64$ and $N=256$.} \label{Figex3}
\end{figure}

\begin{example}{\it Unbalanced channels}:
\label{ex3} Consider example \ref{ex2} with power unbalance and/or
channel loss in transmission system, i.e. the true model at stage
$s$ is
\begin{equation}
\label{ve7} r(n)=\sum\limits_{m=1}^{M}\beta_m
w^s_m\alpha^{(s-1)}_m c_m(n)+v(n),
\end{equation}
where $0<\beta_m\leq 1$ for all $1\leq m \leq M$. Both the LMS-PPIC
and the PLMS-PPIC methods assume the model (\ref{e7}), and their
estimations are based on observations $\{r(n),X^s(n)\}$, instead of
$\{r(n),\mathbf{G}X^s(n)\}$, where the channel gain matrix is
$\mathbf{G}=\mbox{diag}(\beta_1,\beta_2,\cdots,\beta_m)$. In this
case we repeat example \ref{ex2}. We randomly get each element of
$G$ from $(0,0.3]$. Results are given in Figure~\ref{Figex3}. As it
is shown, in all cases the PLMS-PPIC method outperforms both the
conventional and the LMS-PPIC methods.
\end{example}

\begin{figure} \centering
\includegraphics[width=.4\textwidth]{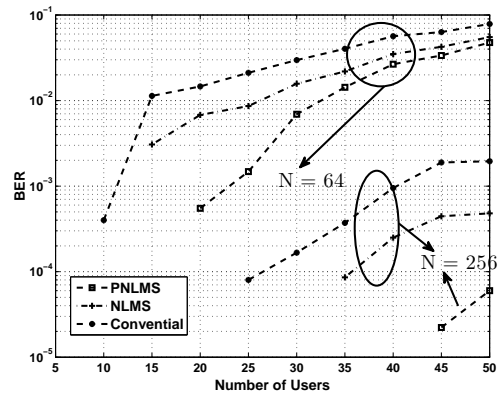}
\caption{The BER of the conventional, the LMS-PPIC, and the
PLMS-PPIC methods versus the system load in time varying Rayleigh
fading channel using two stages for $N=64$ and
$N=256$.}\label{Figex4}
\end{figure}

\begin{example}
\label{ex4} {\it Time varying channels}: Consider example \ref{ex2}
with time varying Rayleigh fading channels. In this case we assume
the maximum Doppler shift of $40$HZ, the three-tap
frequency-selective channel with delay vector of $\{2\times
10^{-6},2.5\times 10^{-6},3\times 10^{-6}\}$sec and gain vector of
$\{-5,-3,-10\}$dB. Results are given in Figure~\ref{Figex4}. As it
is seen while the PLMS-PPIC outperforms the conventional and the
LMS-PPIC methods when the number of users is less than $30$, all
three methods have the same performance when the number of users is
greater than $30$.
\end{example}

\section{Conclusion}\label{S6}%

In this paper, parallel interference cancelation using adaptive
multistage structure and employing a set of NLMS algorithms with
different step-sizes is proposed. According to the proposed method,
in each iteration the parameter estimate is chosen in a way that its
corresponding algorithm has the best compatibility with the true
parameter. Because the step-sizes of all algorithms are chosen from
a stable range, the total system is therefore stable. Simulation
results show that the new method has a remarkable performance for
different scenarios including Rayleigh fading channels even if the
channel is unbalanced.

\begin{table}
\centering{\caption{The procedure of the LMS-PPIC method}
\label{tab1} { \small
\begin{tabular}{|l|l|l|l|l|l|}
\hline \multirow{4}{*}{\rotatebox{90}{Initial Values~~}} &
{$\mbox{for}~~m=1,2,\cdots,M $}
&\multicolumn{4}{|l|}{$\alpha^{0}_m=\mbox{sign}\left\{\mbox{real}\left\{\sum\limits_{n=1}^{N}r(n)c^*_m(n)\right\}\right\}$ } \\
&&\multicolumn{4}{|l|}{} \\
&&\multicolumn{4}{|l|}{} \\
&&\multicolumn{4}{|c|}{} \\
&&\multicolumn{4}{|c|}{} \\
\hline
\multicolumn{2}{|l|}{$\mbox{for}~~s=1,2,\cdots,S $} &\multicolumn{4}{|l|}{$W^{s}(0)=[w^s_1(0),\cdots,w^s_M(0)]^T=[0,\cdots,0]^T$}\\
\cline{3-6} \multicolumn{2}{|l|}{}
&\multirow{4}{*}{\rotatebox{90}{NLMS algorithm~~~}}&{$\mbox{for}~~n=1,2,\cdots,N$}&\multicolumn{2}{|l|}{$X^{s}(n)=[\alpha^{(s-1)}_1c_1(n),\alpha^{(s-1)}_2c_2(n),\cdots,\alpha^{(s-1)}_Mc_M(n)]^T$}\\
\multicolumn{2}{|l|}{} &&&\multicolumn{2}{|l|}{$e(n)=r(n)-W^{s^T}(n-1)X^{s}(n)$} \\
\multicolumn{2}{|l|}{} &&&\multicolumn{2}{|l|}{$Z(n)=\frac{X^{s^*}(n)}{\|X^s(n)\|^2}e(n)$} \\
\multicolumn{2}{|l|}{} &&&\multicolumn{2}{|l|}{$W^{s}(n)=W^{s}(n-1)+\mu Z(n)$} \\
\multicolumn{2}{|l|}{} &&&\multicolumn{2}{|l|}{} \\
\multicolumn{2}{|l|}{} &&&\multicolumn{2}{|l|}{} \\
\cline{3-6} \multicolumn{2}{|l|}{}&
\multicolumn{2}{|l|}{$\mbox{for}~~m=1,2,\cdots,M$}&\multicolumn{2}{|l|}{$q^{s}_{m}(n)=r(n)-\sum\limits_{m^{'}=1,m^{'}\ne m}^{M}w^{s}_{m^{'}}(N)\alpha^{(s-1)}_{m^{'}} c_{m^{'}}(n)$} \\
\multicolumn{2}{|l|}{}& \multicolumn{2}{|l|}{}&\multicolumn{2}{|l|}{$\alpha^{s}_m=\mbox{sign}\left\{\mbox{real}\left\{\sum\limits_{n=1}^{N}q^s_m(n)c^*_m(n)\right\}\right\}$} \\
\hline
\end{tabular} }}
\end{table}

\begin{table}
\caption{The procedure of the PLMS-PPIC method} \label{tab2}
\begin{center}{{ \small
\begin{tabular}{|l|l|l|l|l|l|}
\hline \multirow{4}{*}{\rotatebox{90}{Initial Values~~}} &
{$\mbox{for}~~m=1,2,\cdots,M $}
&\multicolumn{4}{|l|}{$\alpha^{0}_m=\mbox{sign}\left\{\mbox{real}\left\{\sum\limits_{n=1}^{N}r(n)c^*_m(n)\right\}\right\}$ } \\
&&\multicolumn{4}{|l|}{} \\
&&\multicolumn{4}{|l|}{} \\
&&\multicolumn{4}{|c|}{} \\
&&\multicolumn{4}{|c|}{} \\
\hline
\multicolumn{2}{|l|}{$\mbox{for}~~s=1,2,\cdots,S $} &\multicolumn{4}{|l|}{$W^{s}(0)=[w^s_1(0),\cdots,w^s_M(0)]^T=[0,\cdots,0]^T$}\\
\cline{3-6} \multicolumn{2}{|l|}{}
&\multirow{4}{*}{\rotatebox{90}{PNLMS algorithm ~~~~~~~~~}}&{$\mbox{for}~~n=1,2,\cdots,N$}&\multicolumn{2}{|l|}{$X^{s}(n)=[\alpha^{(s-1)}_1c_1(n),\alpha^{(s-1)}_2c_2(n),\cdots,\alpha^{(s-1)}_Mc_M(n)]^T$}\\
\multicolumn{2}{|l|}{} &&&\multicolumn{2}{|l|}{$e(n)=r(n)-W^{s^T}(n-1)X^{s}(n)$} \\
\multicolumn{2}{|l|}{} &&&\multicolumn{2}{|l|}{$Z(n)=\frac{X^{s^*}(n)}{\|X^s(n)\|^2}e(n)$} \\
\multicolumn{2}{|l|}{} &&&\multicolumn{2}{|l|}{$\mbox{min}=\infty, l=1$} \\
\cline{5-6}
\multicolumn{2}{|l|}{} &&&$\mbox{for}~~k=1,2,\cdots,L$ &$W^{s}_k(n)=W^{s}(n-1)+\mu_{k}Z(n)$ \\
\multicolumn{2}{|l|}{} &&&&$\mbox{if}~~\sum\limits_{m=1}^{M}||w^s_{m,k}(n)|-1|<\mbox{min}:$ \\
\multicolumn{2}{|l|}{} &&&&$~~~~~\mbox{min}=\sum\limits_{m=1}^{M}||w^s_{m,k}(n)|-1|$\\
\multicolumn{2}{|l|}{} &&&&$~~~~~l=k$\\
\cline{5-6}
\multicolumn{2}{|l|}{} &&&\multicolumn{2}{|l|}{$W^s(n)=W^s_l(n)$}\\
\cline{3-6}
\multicolumn{2}{|l|}{}&\multicolumn{2}{|l|}{$\mbox{for}~~m=1,2,\cdots,M$}&\multicolumn{2}{|l|}{$q^{s}_{m}(n)=r(n)-\sum\limits_{m^{'}=1,m^{'}\ne m}^{M}w^{s}_{m^{'}}(N)\alpha^{(s-1)}_{m^{'}} c_{m^{'}}(n)$} \\
\multicolumn{2}{|l|}{}& \multicolumn{2}{|l|}{}&\multicolumn{2}{|l|}{$\alpha^{s}_m=\mbox{sign}\left\{\mbox{real}\left\{\sum\limits_{n=1}^{N}q^s_m(n)c^*_m(n)\right\}\right\}$} \\
\hline
\end{tabular} }}
\end{center}
\end{table}

\end{document}